\documentclass[universe,article,accept,moreauthors,pdftex,12pt,a4paper]{mdpi}
\lastpage{x}
\doinum{10.3390/------}
\pubvolume{1}
\pubyear{2015}
\history{Received: 15 October 2014  / Accepted: 28 January 2015  / Published: xx}
\usepackage{graphicx}
\usepackage{amssymb}
\usepackage{hyperref}
\usepackage{natbib}
\usepackage{booktabs}
\usepackage{multirow}
\usepackage{soul}
\newcommand{\be}{\begin{equation}}
\newcommand{\ee}{\end{equation}}
\newcommand{\bq}{\begin{eqnarray}}
\newcommand{\eq}{\end{eqnarray}}

\Title{Observational Constraints on Varying-Alpha Domain Walls}

\Author{Pedro P. Avelino $^{1,}$ $^{2,}$*, {Lara Sousa} $^{1}$ $\dagger$}

\address{%
$^{1}$ Instituto de Astrof\'{i}sica e Ci\^encias do Espa\c{c}o, Universidade do Porto, CAUP, Rua das Estrelas, PT4150-762 Porto, Portugal;\\
$^{2}$ Departamento de F\'{\i}sica e Astronomia, Faculdade de Ci\^encias, Universidade do Porto, Rua do Campo Alegre 687, PT4169-007 Porto, Portugal;\\
$\dagger$ E-Mail: Lara.Sousa@astro.up.pt
 \vspace{-12pt}}
\corres{E-Mail: pedro.avelino@astro.up.pt; \vspace{0pt}}

\externaleditor{{Academic Editor: Lorenzo Iorio}\vspace{0pt}}
\abstract{We consider the possibility that current hints of spatial variations of the fine structure constant at high redshift, based on VLT/UVES and Keck/HIRES observations, could be caused by a biased domain wall network described by a scalar field non-minimally coupled to the electromagnetic field. We show that in order to be responsible for the reported spatial variations of the fine structure constant, the fractional contribution of the domain wall network to the energy density of the Universe should be tightly constrained within the range $10^{-10} < \Omega_{w0} < 10^{-5}$. We also show that the domain wall dynamics should be essentially frictionless, so that its characteristic scale is in the order of the Hubble radius at the \mbox{present time}}


\begin{document}

%%%%%%%%%%%%%%%%%%%%%%%%%%%%%%%%%%%%%%%%%%
\vspace{-12pt}
\section{Introduction}

Recent analysis of a combined sample of quasar absorption line spectra obtained using UVES (the Ultraviolet and Visual Echelle Spectrograph) on the Very Large Telescope (VLT) and HIRES (the High Resolution Echelle Spectrometer) on the Keck Telescope have provided hints of a spatial variation of the fine structure constant ($\alpha$) that is well represented by an angular dipole model \cite{Webb:2010hc,King:2012id}. \mbox{Subsequently, several} authors have shown that these results could be induced by a domain wall network described by a scalar field non-minimally coupled to the electromagnetic field \cite{Olive:2010vh, Chiba:2011en, Bamba:2011nm,Olive:2012ck} (see also \cite{Menezes:2004tp} for the first discussion of the cosmological implications of varying-$\alpha$ walls).

Domain wall networks tend to dominate the energy density of the universe at late times, a property shared by all dark energy candidates. This means that, unlike cosmic strings, domain walls are not expected to lead to observable signatures on cosmic structure formation and to contribute significantly to the primary Cosmic Microwave Background (CMB) anisotropies. Although frozen domain wall networks have been proposed in the past as a dark energy candidate \cite{Bucher:1998mh}, detailed studies have since ruled out any significant contribution of domain walls to the dark energy budget \cite{PinaAvelino:2006ia,Avelino:2008ve} \mbox{(see also \cite{Sousa:2009is,Avelino:2010qf,Sousa:2011iu})}.

As is the case with other topological defects, considering that a small domain wall contribution to the energy budget cannot be eliminated by cosmological observations, domain walls may still play a significant cosmological role. In this paper, we study the potential role of a domain wall network as a source of the reported variations of $\alpha$ based on VLT/UVES and Keck/HIRES observations, and derive tight observational constraints on this scenario.

\section{Domain Wall Evolution: The Basics}

In this section, we shall briefly review the essential aspects driving the cosmological evolution of a domain wall. In a flat homogeneous and isotropic Friedmann--Robertson--Walker (FRW) background, the line element is given by
\be ds^2=-dt^2+a^2(t)\left[dr^2+r^2\left(d\theta^2 +\sin^2 \theta d\phi^2\right)\right]\,
\ee where $t$ is the physical time, $a$ is the scale factor and $(r,\theta,\phi)$ are spherical coordinates. In such a background, the dynamics of a domain wall is essentially determined by two competing factors: the acceleration caused by domain wall curvature and the Hubble damping. The evolution equation of a domain wall in FRW universe may be rigorously derived from the Nambu--Goto action (see \citep{Sousa:2011iu}). \mbox{For the} purpose of this paper, however, it is sufficient (and considerably more succinct) to study these effects individually.

Let us start by considering the case of a spherically symmetric domain wall in a Minkowski spacetime (with $a=1$). This will allow us to pinpoint the effect of curvature on domain wall dynamics. In this case, the wall has an invariant area, $S=4 \pi \gamma r^2$ (where $\gamma \equiv (1-v^2)^{-1/2}$, and $v$ is the domain wall velocity), that is proportional to its energy (this is the domain wall analogue of the invariant perimeter of a cosmic string loop, $2\pi \gamma r$). It is straightforward to show that energy conservation then leads to the following equation of motion for a spherical domain wall in a Minkowski spacetime
\be
\frac{dv}{dt}=n(1-v^2)\frac{f_k}{r}\,
\label{v-curv}
\ee where $n=2$ and $f_k$ can be equal to $1$ or $-1$ depending, respectively, on whether the curvature is accelerating ($dv/dt>0$) or decelerating ($dv/dt<0$) the domain wall. Note that this expression also applies to point particles, if one sets $n=0$, and to cosmic strings, for $n=1$.

Note however that, in an FRW universe, the motion of the domain walls will be damped as a result of the expansion of the background, and consequently domain wall energy is no longer conserved. In such a background, the wall momentum per unit comoving area should be conserved. One thus has that, for a planar wall
\be
\frac {d(v\gamma a^{n+1})}{dt}=0\,
\ee or, alternatively
\be
\frac {dv}{dt}=-(1-v^2)(n+1)Hv\,
\label{v-Hubble}
\ee with $n=0$, $1$ and $2$, for point particles, cosmic strings and domain walls, respectively. Here, \mbox{ $H=(da/dt)/a$} is the Hubble parameter.

Equations (\ref{v-curv}) and (\ref{v-Hubble}) may be combined in order to obtain the equation of motion of a spherically symmetric domain wall in an FRW background. Although this equation describes the microscopic evolution of a single spherical wall, it also captures the essential features of the large scale dynamics of domain wall networks. As a matter of fact, a close analogue has been shown to accurately describe the root-mean-square velocity of domain wall networks. The reader may visit references~\cite{PinaAvelino:2006ia,Avelino:2008ve,Sousa:2010zz,Avelino:2010qf,Avelino:2011ev,Sousa:2011ew,Sousa:2011iu} for a detailed description of the semi-analytical Velocity-dependent One-Scale (VOS) model for domain \mbox{wall networks}.

\section{The Biased Evolution of Varying-$\alpha$ Walls}

For the remainder of this paper, we shall assume that the domain walls are associated with the variation of a scalar field $\phi$ that interpolates from $\phi_-$ to $\phi_+$ (or vice versa) at the domain wall. Note that this is the simplest realisation of a domain wall network without junctions and that more complex networks with junctions may be constructed in models with more than two minima of the scalar field \mbox{potential \cite{PinaAvelino:2006ia,Avelino:2008ve}}. Let us also assume that the value of the fine structure constant $\alpha$ is a function of the scalar field $\phi$ ($\alpha=\alpha(\phi)$)---so that it takes different values on each of the wall's domains---and let $\alpha_-\equiv\alpha(\phi_-)$ and $\alpha_+\equiv\alpha(\phi_+)$ be the values of $\alpha$ on each side of the domain wall. Due to the (weak) dependence of the quark masses on $\alpha$, the baryons on opposite sides of a domain wall have slightly different masses. \mbox{The fractional} mass difference is given approximately by
\be
\zeta \equiv \left|\frac{\Delta m}{m}\right| = \left|\xi \frac{\Delta \alpha}{\alpha}\right|
\ee where $\xi$ corresponds to the fractional electromagnetic contribution to the baryon mass. Note that $\xi$ takes different values depending on the nature of the particles under consideration: $\xi=-0.0007$ and $\xi=0.00015$, for the proton and neutron respectively \cite{Gasser:1982ap}. Protons are significantly more abundant than neutrons and the absolute value of their electromagnetic contribution is considerably larger than that of neutrons. For this reason, and for the remainder of this paper, we shall assume that $\xi=-0.0007$.

The domains on opposite sides of a wall are expected to have the same average baryon number density. However, as a consequence of the difference in baryon mass, they must have different average baryon energy densities. This energy difference between the domains introduces a bias on the dynamics of the domain walls, favouring the domains with a smaller value of the baryon energy density. Biased domain walls were originally envisioned as means to evade the Zel'dovich bound \cite{ZEL}, but they have also served as the basis of the devaluation scenario that attempted (and failed) to solve the cosmological constant problem \cite{Freese:2005pu}. The dynamics of biased domain walls has been studied in detail in these contexts (see, e.g., \cite{Sikivie:1982qv,Gelmini,Larsson:1996sp,Avelino:2008qy,Avelino:2008mh}). Here, we follow closely the approach in \cite{Avelino:2008qy}.

A generic form of a Lagrangian allowing for varying-alpha domain walls would be
\be {\mathcal L}=-\frac12 \partial_\mu \phi \partial^\mu \phi -V_{eff}(\phi)\,
\ee where the effective potential for the scalar field $\phi$ is given by
\be V_{eff}(\phi)=V(\phi)+\rho_B(\phi)\,
\ee and $V(\phi)$ is a scalar field potential allowing for domain wall solutions, such as a standard quartic potential with two degenerate minima
\be V(\phi) = V_0 \left(\frac{\phi^2}{\eta^2}-1\right)^2\,
\ee

Here $\eta=|\phi_\pm|$ and $\rho_B(\phi)$ is the baryon density, which in this case is a function of the scalar field $\phi$ coupled to electromagnetism that drives the space-time variations of $\alpha$. Although the above Lagrangian does not describe in detail microscopic interactions nor takes into account thermal corrections, it provides a generic description of the macroscopic interaction between the baryons and the domain walls.

Let us denote the values of the baryon densities on each side of the domain walls by $\rho_{B-} \equiv \rho_{B}(\phi_-)$ and $\rho_{B+} \equiv \rho_{B}(\phi_+)$ and introduce a bias parameter defined by
\be
\epsilon=\rho_{B+}-\rho_{B-}=\zeta \rho_B\,
\ee where $\rho_B$ is the average baryon energy density. In order to model the effect of this bias on the dynamics of the domain wall, we start by considering a planar domain wall in a Minkowski spacetime (so as to isolate the effect of bias on the dynamics). Energy conservation implies that
\be d(\sigma \gamma)=v\epsilon dt\,
\ee or equivalently
\be
\frac{dv}{dt}=\frac{\epsilon f_\epsilon}{\sigma \gamma^3}\equiv \frac{f_\epsilon}{R_{\epsilon}\gamma^3}\,
\label{v-bias}
\ee where $\sigma$ is the domain wall tension (equal to the domain wall energy per unit area in the wall's rest frame) and $f_\epsilon$ can be equal to $1$ or $-1$ depending on whether the bias is accelerating ($dv/dt>0$) or decelerating ($dv/dt<0$) the domain wall---an alternative way to arrive at Equation (\ref{v-bias}) would be using momentum conservation: $d(\sigma \gamma v)=\epsilon dt$. When a domain wall is moving towards a region with higher baryon energy density, the wall gains momentum to balance the resulting energy loss. Therefore, the domain walls feel a pressure that favours the suppression of the regions with a larger value of the baryon energy density. This backreaction effect on the dynamics of varying-alpha walls is a direct consequence of energy-momentum conservation. In the absence of expansion, energy-momentum conservation would require the decrease (increase) of the baryon energy density to be exactly compensated by an increase (decrease) of the energy density of the domain walls. One may infer from Equation (\ref{v-bias}) that the effect of the energy difference between the domains is to accelerate the domain wall and, in that sense, $\epsilon$ acts as an effective curvature characterised by a constant length scale $R_{\epsilon}=\sigma/\epsilon$.

The effect of the bias on the dynamics of domain walls with curvature in an expanding background is determined by the interplay of the surface pressure (which is caused by curvature), the volume pressure (which results from $\epsilon$) and the damping (which is caused by Hubble expansion). Equations (\ref{v-curv}), (\ref{v-Hubble}) and (\ref{v-bias}) may be combined into the following equation
\be
\frac{dv}{dt}=(1-v^2)\left(\frac{2f_k}{R} + \frac{f_\epsilon}{R_{\epsilon} \gamma} -3Hv\right)\,
\label{v-full}
\ee where $R=ar$ is the physical radius of the domain wall. Biased domain walls may be long-lived or disappear almost immediately, depending on the relative importance of the different terms in \mbox{Equation (\ref{v-full})}. As the domain walls evolve and the physical size of the domains increases, the importance of the bias term, when compared with the curvature and damping terms, grows with cosmic time. \mbox{When the} domains become larger than the bias scale---such that $R>R_{\epsilon}$---the walls decay due to the effect of the volume pressure. As a matter of fact, it has been shown that this result also applies to domain wall networks with junctions \cite{Avelino:2008mh}:
domains with larger baryon energy density disappear exponentially fast  once the characteristic length scale of $L$ becomes larger than $R_{\epsilon}$ \cite{Larsson:1996sp} ---or, equivalently, when
\be
\frac{\sigma}{\epsilon L}=\frac{\rho_{w}}{\epsilon}=\frac{\Omega_w}{\Omega_{B}\zeta} < 1\,
\label{constraint}
\ee
Here $\rho_c$ is the critical density of the universe, $\rho_w=\sigma/L$ is the average energy density of the domain wall network, $\Omega_{B}=\rho_B/\rho_c$ and $\Omega_{w}=\rho_w/\rho_c$. Varying-$\alpha$ domain walls have, thus, an additional source of instability---the bias introduced by the energy difference between domains. As we shall see in the next section, this will allow us derive a lower bound on the admissible varying-$\alpha$ domain wall energy density.

\section{Observational Constraints}

Observations of the quasar absorption spectra at high redshifts obtained using HIRES on the Keck Telescope were shown to be consistent with a fractional variation of $\alpha$ of $(0.57\pm0.11)\times 10^{-5}$ \cite{Murphy:2003hw} \mbox{(see also} \cite{Webb:1998cq,Murphy:2000pz,Webb:2000mn}). However, ensuing studies based on VLT/UVES data have indicated a much smaller value of $\Delta \alpha/\alpha$ \cite{Chand:2004ct,Quast:2003qu}. These observational constraints can be reconciled by allowing for a spatial variation of the fine structure constant. In fact, it has been shown in \cite{King:2012id} that the data from a large sample of 295 absorbers from the VLT and Keck telescopes is compatible with an angular dipole model with amplitude \cite{King:2012id} (see also \cite{Webb:2010hc})
\be
\Delta\alpha/\alpha=0.97^{+0.22}_{-0.20}\times 10^{-5}\,
\label{daloval}
\ee

Domain wall networks described by a scalar field non-minimally coupled to the electromagnetic field have been suggested as a potential source of variations of $\alpha$ \cite{Olive:2010vh, Chiba:2011en, Bamba:2011nm,Olive:2012ck} (see also \cite{Menezes:2004tp}). It has been shown that this scenario is compatible with the current data, but a detailed verification will require a new generation of high-resolution ultra-stable spectrographs such as ESPRESSO (the Echelle SPectrograph for Rocky Exoplanet and Stable Spectroscopic Observations). Note that, in a model in which spatial variations of $\alpha$ are caused by a domain wall network, the stringent laboratory constraints (see, for example, \cite{Uzan:2010pm} and references therein) are naturally evaded.

The domain wall network would have to have a characteristic scale in the order of the Hubble radius in order to be able to induce spatial variations on a cosmological scale. The redshift of the closest domain wall may be estimated by solving the equation

\be
\Delta \eta \equiv \eta_0-\eta = \beta L/a\,
\ee with $\beta \lesssim 1$, using the VOS model for domain walls to estimate the characteristic length of the network. Here $\eta$ is the conformal time (or comoving particle horizon) defined by $\eta=\int dt/a$, and the subscript ``0'' refers to the present time. Using the calibration of the VOS model given in \cite{Leite:2012vn} and assuming fractional matter and cosmological constant energy densities compatible with the Planck results \cite{Ade:2013zuv} ($\Omega_{m0}=0.315$ and $\Omega_{\Lambda 0}=1-\Omega_{m0}$, respectively), we have found that if, as indicated by the results obtained in \cite{Olive:2012ck}, the wall closest to us is at a redshift $z = $ 0.5--1, its existence would be consistent with a plausible range for the corresponding value of $\beta$:

\be
0.3 \lesssim \beta \lesssim 0.7 %\beta=0.3-0.7\,
\ee

Results from high resolution field theory simulations have shown no significant dependence of the characteristic velocity on the properties of the network. However, the characteristic scale of the network is highly dependent on the network configuration, and its value can vary significantly: it may be smaller by a factor of up to $3$ in the case of complex domain wall networks with junctions \cite{Avelino:2008ve}, or even by a larger factor if friction is important \cite{Sousa:2011iu} or in the presence of massive junctions \cite{Sousa:2009is}. The above estimate of the value of $\beta$ appears to favour the simplest frictionless domain wall models without junctions over more complex scenarios with junctions.

A necessary requirement for a domain wall network to be able to seed the spatial variations of $\alpha$ is that the bias introduced by the interaction with the baryons does not lead to the suppression of the domain wall network prior to the present epoch. This requirement (see Equation (\ref{constraint})) yields the following lower limit on the contribution of domain walls to the cosmic energy budget
\be
\Omega_{w0} > \Omega_{B0} \left|\xi\frac{\Delta \alpha}{\alpha}\right| \sim 3 \times 10^{-5} \left|\frac{\Delta \alpha}{\alpha}\right|\,
\label{lowerb}
\ee where we have taken $\Omega_{B0}h^2=0.02205$, with $h=0.673$, as indicated by Planck data \cite{Ade:2013zuv}, \mbox{and $\xi=-0.0007$}. Assuming that the reported spatial variations of $\alpha$, based on observations from VLT/UVES and Keck/HIRES spectrographs, are due to varying-$\alpha$ walls and using the constraint on the $\alpha$-variation given by Equation (\ref{daloval}), one finds that
\be
\Omega_{w0} > 10^{-10}\,
\label{lowerb1}
\ee

On the other hand, a conservative constraint on the contribution of a domain wall network with a characteristic length in the order of the Hubble radius to the cosmic energy budget is
\be
\Omega_{w0} < 10^{-5}\,
\label{omup}
\ee in order to prevent domain walls from providing the dominant contribution to the large scale temperature anisotropies of the CMB. Note that this upper bound is conservative and that a more detailed analysis, using the current CMB constraints from Planck and the Wilkinson Microwave Anisotropy Probe (WMAP), would most certainly lead to a tighter upper bound.

Hence, we have found that the observational window for domain wall networks as a source of the reported spatial variations of $\alpha$ is narrow:
\be 10^{-10} < \Omega_{w0} < 10^{-5}\,
\ee

The scenario of domain-wall-induced spatial variations of $\alpha$ is thus tightly constrained.

\section{\label{conc}Discussion and Conclusions}

In this paper, we have considered the possibility that a domain wall network may be responsible for the reported spatial variations of $\alpha$ based on observations from VLT/UVES and Keck/HIRES spectrographs. We have shown that in order to explain these variations, the dynamics of the domain wall network needs to be essentially frictionless (so that its characteristic length scale at the present time is in the order of the Hubble radius) and its fractional energy density should be within five orders of magnitude \mbox{($10^{-10} < \Omega_{w0} < 10^{-5}$)}. The presence of a domain wall network with a characteristic scale comparable to the Hubble radius may leave a variety of other observational signatures---in particular if its fractional energy density is sufficiently close to the upper limit set in Equation (\ref{omup}). For instance, as happens in the case of cosmic strings \cite{Seljak:2006hi,Pogosian:2007gi,Bevis:2007qz,Sanidas:2012ee,Kuroyanagi:2012wm,Sousa:2013aaa,Sousa:2014gka}, as domain wall networks evolve and interact, a fraction of their energy is released in the form of vector and tensor modes \cite{Hiramatsu:2013qaa}, which may contribute to the B-mode polarisation of the CMB.

Recent results by the Background Imaging of Cosmic Extragalactic Polarization (BICEP2) experiment seem to be consistent with an excess of B-mode power on large angular scales (in the multipole range $30 < l < 150$), which has been interpreted as the first direct evidence of a primordial gravitational wave background produced at an early inflationary stage at the Grand Unified Theory\mbox{ scale \cite{Ade:2014xna}}. However, in \cite{Lizarraga:2014eaa,Moss:2014cra,Lizarraga:2014xza}, alternative and/or complementary interpretations involving cosmic strings and other cosmic defects have been proposed. For sufficiently large values of $\Omega_{w0}$ ($\sim$$10^{-6}-10^{-5}$) the domain walls are expected to provide a significant contribution to the B-mode polarisation power spectrum associated to vector and tensor perturbations, thus being a possible explanation of the B-mode polarisation signature detected by BICEP2 \cite{Ade:2014xna}. In \cite{Moss:2014cra} it was found that, in order for cosmic strings to be the dominant contributor to excess of B-mode power on large angular scales detected by BICEP2, the inter-string distance would need to be extremely large. Although this analysis does not apply directly to domain wall networks, the fact that the characteristic length of a domain wall network is significantly larger than that of local strings is interesting. Furthermore, since domain walls only become cosmologically relevant at recent times, B-mode polarisation induced by domain walls should be strongly suppressed on small scales when compared with a corresponding contribution from cosmic strings, despite the fact that it is expected to be comparable to that of strings at low $l$, for the same average defect energy density at the present time (for this reason domain walls, unlike other defect models, provide a negligible contribution to the matter power spectrum and to the primary CMB anisotropies). These facts could make domain walls a stronger contender than other defects (such as cosmic strings) to explain the excess of B-mode power detected by BICEP2 on large angular scales.

Another interesting observation is the apparent tension between the original interpretation of the BICEP2 results and the Planck upper limit on the tensor-to-scalar ratio $r$. A recent suggestion to solve this discrepancy involves a spatial variation of $r$ that could also account for the large scale anomalies in the temperature distribution of the cosmic microwave background detected by Planck and \mbox{WMAP \cite{Ade:2013zuv,Ade:2013ktc,Hinshaw:2012aka,Bennett:2012zja,Chluba:2014uba}}. A similar signature would also be expected in the case of the domain wall scenario studied in the present paper. The large characteristic length scale of the domain wall network at the present time could be naturally associated with large scale temperature and polarisation CMB power asymmetries and with a large cosmic variance that future studies will need to tackle.

The origin of the signal detected by BICEP2 is still a matter of controversy, with several authors questioning the validity of the methods used (see, e.g., \citep{Mortonson:2014bja,Flauger:2014qra}). Still, even if the cosmological origin of the signal is ruled out, stronger upper bounds on the fractional energy density of domain walls may be inferred. In either case, valuable information might be gained by performing more detailed studies of the large scale anisotropies and CMB polarisation signatures generated by varying-$\alpha$ domain walls. Such studies will require an accurate computation of the scale dependence of both vector and tensor perturbations generated by domain wall networks around the present time for a wide range of representative domain wall network models (with or without junctions), and will be the subject of \mbox{future work}.
%%%%%%%%%%%%%%%%%%%%%%%%%%%%%%%%%%%%%%%%%%

\acknowledgments{Acknowledgments}

P.P.A. is supported by Funda{\c c}\~ao para a Ci\^encia e a Tecnologia (FCT) through the Investigador FCT contract of reference IF/00863/2012 and POPH/FSE (EC) by FEDER funding through the program ``Programa Operacional de Factores de Competitividade-COMPETE''. L.S. is supported by Funda\c{c}\~{a}o para a Ci\^{e}ncia e Tecnologia (FCT, Portugal) and by the European Social Fund (POPH/FSE) through the grant SFRH/BPD/76324/2011. This work was also partially supported by grant \mbox{PTDC/FIS/111725/2009 (FCT)}.

%%%%%%%%%%%%%%%%%%%%%%%%%%%%%%%%%%%%%%%%%%
\authorcontributions{Author Contributions}

The authors contribute equally to this paper. All authors have read and approved the final version.

%%%%%%%%%%%%%%%%%%%%%%%%%%%%%%%%%%%%%%%%%%

\conflictofinterests{Conflicts of Interest}

The authors declare no conflict of interest.

%=================================================================
% References: Variant A
%=================================================================
% Back Matter (References and Notes)
%----------------------------------------------------------
% Style and layout of the references

%\makeatletter
%\renewcommand\@biblabel[1]{#1. }
%\makeatother
%\bibliography{alpha_wall}
%\bibliographystyle{mdpi}

%\end{thebibliography}

%=================================================================
% References: Variant B
%=================================================================
% Use the following option to include external BibTeX files:
%\bibliography{lite}
%\bibliographystyle{mdpi}

%%%%%%%%%%%%%%%%%%%%%%%%%%%%%%%%%%%%%%%%%%

%\abbreviations{Abbreviations/Nomenclature}
%
%Main text.

%%%%%%%%%%%%%%%%%%%%%%%%%%%%%%%%%%%%%%%%%%

%\appendix
%\section{Appendix Title}
%
%Main text.

\end{document}